\documentclass[12pt]{article}
\usepackage{graphicx}

\newcommand\pubnumber{eConf C1304143 }
\newcommand\pubdate{\today}

\newcommand{\msun}{\mbox{$M_\odot$}}

\def\be{\begin{equation}}
\def\ee{\end{equation}}
\def\bary{\begin{eqnarray}}
\def\eary{\end{eqnarray}}

\def\napoli{Instituto de Astronom\'ia \\
Universidad Nacional Autonoma de M\'exico, CU M\'exico}
\def\support{\footnote{Luc Binette-Fundaci\'on UNAM Fellow.}}

\def\Title#1{\begin{center} {\Large #1 } \end{center}}
\def\Author#1{\begin{center}{ \sc #1} \end{center}}
\def\Address#1{\begin{center}{ \it #1} \end{center}}

\newcommand\pubblock{\rightline{\begin{tabular}{l} \pubnumber\\
         \pubdate  \end{tabular}}}
\newenvironment{Abstract}{\begin{quotation}  }{\end{quotation}}
\newenvironment{Presented}{\begin{quotation} \begin{center} 
             PRESENTED AT\end{center}\bigskip 
      \begin{center}\begin{large}}{\end{large}\end{center} \end{quotation}}
\def\Acknowledgements{\bigskip  \bigskip \begin{center} \begin{large}
             \bf ACKNOWLEDGEMENTS \end{large}\end{center}}

\begin{document}
\begin{titlepage}
\pubblock

\vfill
\Title{Neutrino Oscillation as a constraint in the dynamics of  Pop III Gamma Ray Bursts}
\vfill
\Author{Nissim Fraija\support and Enrique Moreno M\'endez}
\Address{\napoli}
\vfill
\begin{Abstract}
Population III stars are believed to be rapidly-rotating sources with a mass range of hundreds to thousand of solar masses.  Masses larger than  260 M$_\odot$  are expected to collapse resulting in central rotating Kerr black holes with large rotation parameter a.  Due to particle-antiparticle asymmetry  is small, we use the  neutrino effective potential up to order $1/M_W^4$ in a magnetized plasma at the base of the ejecta  to constrain the rotation parameter by means of  neutrino oscillations.  Additionally, we investigate the implications in the magnetic field, temperature and electron asymmetry.
\end{Abstract}
\vfill
\begin{Presented}
2013 GRB Symposium of GRBs\\
Nashville, Tennessee,  April 14--18, 2013
\end{Presented}
\vfill
\end{titlepage}
\def\thefootnote{\fnsymbol{footnote}}
\setcounter{footnote}{0}

\section{Introduction}

Population III (Pop III) stars are widely believed to have formed at z $\geq 20$ and to be rapidly-rotating stars with a mass range of hundreds of solar masses.  
In particular, those with mass 140  $M_\odot$ $<$ $M_\star$ $<$ 260 $M_\odot$ are expected to explode as pair-instability supernovae and entirely release  their material to the intergalactic medium.  While stars with $M_\star \geq$ 260 $M_\odot$  are expected to collapse resulting in black holes (BHs) with very little mass loss \cite{heg02,fry01}.  
Because of the very high mass involved and its expected fast rotation during gravitation core collapse (GCC), accretion disks and subsequently ultra-relativistic jets are expected to develop, giving rise to highly energetic gamma-ray bursts (GRBs) \cite{heg03, kom09, kom10}.\\
Within the base of jet, the inverse-beta decay of protons ($p +e \to n+  \nu_e$), electron-positron annihilation ($e^++e^-\to Z\to \nu_j+\bar{\nu}_j$ for $j=e,\nu,\tau$) and nucleon-nucleon bremsstrahlung ($NN\to NN+\nu_j+\bar{\nu}_j$, $j=e,\nu,\tau$)   will produce thermal neutrinos which propagate in it. Also neutrinos of similar energies are produced in the accretion disk during the GCC.     Given that electron neutrinos ($\nu_e$) interact with electrons via both, neutral and charged currents (CC), while muon ($\nu_\mu$) and tau ($\nu_\tau)$ neutrinos interact only via the neutral current (NC), their properties will be modified differently when they propagate in a magnetized medium.  The resonant conversion of neutrino from one flavor to another due to the medium effect, is well known as the Mikheyev-Smirnov-Wolfenstein effect \cite{wol78}.  These  neutrino oscillations  have widely been studied in the literature in different scenarios \cite{vol99,das08,jan12,goo86,ruf99}, even though in the fireball environment with weak magnetic field \cite{sah05, sah09a, sah09b} but not with high magnetic field.
\section{Dynamics of Poynting-Dominated  jet}
As a result of the  GCC,  there is a disk of outer radius of $R_d=R_\star/4$ (where $R_\star$ is the stellar radius), disk mass $M_d$, and a central Kerr BH of mass $M_h$ and spin parameter $a_\star = Jc/GM_h^2$ (with $J$ the angular momentum of the BH).  
The  mean accretion rate is $\dot{M}\sim \frac{M_d}{t_{ac}}$, where t$_{ac}\sim(14/9\alpha)(R_\star^3/GM)^{1/2}$ is the accretion time and $\alpha$ is a disk viscosity parameter. When the jet is just developing, the base of the jet is at
\be
r=G\,M_h\,f_1(a_\star)
\ee
where G is the gravitational constant and $f_1(a_\star)=(2-a_\star)+2\sqrt{1-a_\star}$.
Here we ask for the spin parameter to be $0.5\leq a_\star \leq 1.0$ in order to have enough angular momentum to form a disk and enough rotational energy to launch a jet capable of drilling its way out of the star.   
The Poynting jet luminosity, $L=\pi/(48\beta)f_1^{3/2}f_2^2 G^{1/2}M_h^{3/2} R_\star$, is related to the comoving magnetic-field as $L/4\pi r_l^2\Gamma^2\sim U_B$. Hence the comoving magnetic field strength  at the base of the jet is given by,
\be
B=\biggl( \frac{\pi}{24\,G^{3/2}}  \biggr)^{1/2}\,M_h^{-1/4}\,R_\star^{-3/4}\,\Gamma^{-1} \beta^{-1/2}\,f_1^{-1/4}(a_\star)\,f_2(a_\star),
\ee
where $\beta$ is the magnetization parameter and $f_2(a_\star)=a_\star/2(1+\sqrt{1-a_\star^2})$.\\
The base of the jet is initially magnetic, involving pairs (e$^\pm$), photons and also leptons and some baryons; hence the dynamics of the jet  are determined by  the temperature, amount of baryons and lepton asymmetry. The comoving pair temperature of the flow is estimated as,
\be
T=\biggl(\frac{1}{192\,G^{3/2}\,\sigma_B\,}\biggr)^{1/4}\,M_h^{-1/8}\,R_\star^{3/8}\,\Gamma^{-3/4} \beta^{-1/4}\,f_1^{-1/8}(a_\star)\,f_2^{1/2}(a_\star).
\ee
At the beginning, the optical depth at the base of the flow is so high that the dynamics are  dominated by energy, entropy and particle energy conservation (\cite{tom11}).

\section{Effective potential at the base of the jet}

The neutrino properties become modified as they  travel through  a magnetized medium and a heat bath.  A massless neutrino acquires an effective mass and an effective potential in the medium. Although neutrinos can not couple directly to the magnetic field, their effect can be felt through coupling to charged particles in the background.  In all the astrophysical and cosmological environments,  the magnetic field is entangled intrinsically with matter and it also affects the particle properties. The effective potential of a particle  was calculated  using field theory formalism \cite{sch51},    from the real part of its self energy diagram. The effective potential  can be written  as \cite{sah09a, sah09b, fra14a},
\be
V_{eff,B}= \sqrt2 G_F \frac{m^3_e}{\pi^2}
\biggl [\Phi_A-\frac{2 m_e E_{\nu}}{M^2_W} \Phi_B \biggr ],
\ee
where the functions $\Phi_A$ and $\Phi_B$ are defined as
\bary
\Phi_A&=&\sum_{l=0}^{\infty} (-1)^l
\sinh\, \alpha
\biggl [
\biggl(1+\frac32\frac{m_e^2}{M_W^2}
-\frac{eB}{M_W^2}\biggr)
\biggl(
 \frac{2}{\sigma} K_2(\sigma) - \frac{B}{B_c} K_1(\sigma)
\biggr)\nonumber\\
&&-\frac{B}{B_c}\biggl(1+\frac{m_e^2}{2 M_W^2}-\frac{eB}{M_W^2} \biggr )
K_1(\sigma)
\biggr ],
\label{fphiA}
\eary
and
\be
\Phi_B =
 \sum_{l=0}^{\infty} (-1)^l
\cosh\, \alpha
\biggl [
\biggl(\frac{8}{\sigma^2}-\frac52 \frac{B}{B_c} \biggr )K_0(\sigma)
+\biggl( 2-4 \frac{B}{B_c} + \frac{16}{\sigma^2} \biggr )\frac{K_1(\sigma)}{\sigma}
\biggr ],
\label{fphiB}
\ee
where the critical magnetic field is  $B_c=m^2/e\simeq4.1\times10^{13}$ G,   $K_i$ is the modified Bessel function of integral order i, $\alpha=\beta\mu(l+1)$ and $ \sigma=\beta m_e(l+1)$.
The electron asymmetry in the background, $L_e=(N_e-\bar{N}_e)/N_\gamma$, is defined through  the photon number density, $N_\gamma=\frac{2}{\pi}\zeta(3)\,T^3$, so  it can be expressed as
\bary
L_e=&&\frac{m_e^3}{2\,\zeta(3)}\,(192\,G^{3/2})^{3/4}\,M_h^{3/8}\,R_\star^{9/8}\,\Gamma^{9/4} \beta^{3/4}\,f_1^{3/8}(a_\star)\,f_2^{-3/2}(a_\star)\cr
&& \times \sum_{l=0}^{\infty} (-1)^l \sinh{\alpha} \left[ \frac{2}{\sigma} K_2(\sigma)-\frac{B}{B_c} K_1(\sigma) \right],
\eary
where,
\be
N_e-{\bar N}_{e}=\frac{m^3}{\pi^2} \sum_{l=0}^{\infty} (-1)^l \sinh{\alpha}
\left[
\frac{2}{\sigma} K_2(\sigma)-\frac{B}{B_c} K_1(\sigma)
\right].
\ee
In an upcoming paper (Fraija \& Moreno M\'endez in progress) we will address the baryon load issue as well as leptonic asymmetry and resonance length.
\section{Neutrino Mixing}
To determine the neutrino-oscillation probabilities we have to solve the evolution equation of the neutrino system in matter. In the two and  three-flavor framework, this equation is given by \cite{fra14b}
\be
U\cdot \frac{1}{2E_\nu} \textbf{M}\cdot  U^\dagger + diag(V_{eff},\vec{0})
\ee
where
\begin{equation}\label{p1}
\textbf{M}=
\cases{
(-\delta m^2, 0)			 			&  for two flavors, \cr
(-\delta m^2_{21}, 0,\delta m^2_{32})        &  for three flavors.\cr
}
\end{equation}
Here  $\delta m^2$ is the mass difference, $V_{eff}$ is the potential difference between $V_{\nu_e}$ and $V_{\nu_{\mu, \tau}}$,    $E_{\nu}$ is the neutrino energy and $\theta$ is the neutrino mixing angle.  Applying the resonance condition given by,
\begin{equation}\label{recond}
2 \times 10^6  E_\nu V_{eff} =
\cases{
\delta m^2 \cos 2\theta, 		&  for two flavors \cr
\delta m_{32}^2 \cos 2\theta_{13},   &  for three flavors \cr
}
\end{equation}
we obtain that the resonance length which can be written as
\begin{equation}\label{releng}
L_{res}= 4\pi E_\nu
\cases{
\frac{1}{\sqrt{(2 E_\nu V_{eff} -   \delta m^2 \cos 2\theta  )^2+ (\delta m^2 \sin 2\theta)^2}}, 		&  for two flavors, \cr
\frac{1}{\sqrt{(2 E_\nu V_{eff} -   \delta m_{32}^2 \cos 2\theta_{13}  )^2+ (\delta m_{32}^2 \sin 2\theta_{13})^2}},   &  for three flavors. \cr
}
\end{equation}
Considering the adiabatic condition at the resonance, we  can express it as
\begin{equation}\label{p2}
K_{res} =\frac{1}{2\pi E^2_\nu}\,\biggl(\frac{d V_{eff}}{d r}\biggr)^{-1}
\cases{
\delta m^2 \sin 2\theta, 		&  for two flavors \cr
\delta m_{32}^2 \sin 2\theta_{13},   &  for three flavors \cr
}
\end{equation}

\begin{table}[t]
\begin{center}\renewcommand{\tabcolsep}{0.7cm}
\renewcommand{\arraystretch}{1.25}%
\begin{tabular}{l|ccc}\hline
\small{Experiment}   & \small{Mass diference} &\small{Angle}  \\ \hline \hline

\small{Solar\,neutrino \cite{aha11}}&{\small $\delta m^2=(5.6^{+1.9}_{-1.4})\times 10^{-5}\,{\rm eV^2}$}&{\small $\tan^2\theta=0.427^{+0.033}_{-0.029}$}        \\
\small{Atmospheric \,neutrino \cite{abe11a}}&{\small $\delta m^2=(2.1^{+0.9}_{-0.4})\times 10^{-3}\,{\rm eV^2}$}   & {\small $\sin^22\theta=1.0^{+0.00}_{-0.07}$}     \\
\small{Accelerator\,neutrino\cite{chu02}}&{\small $\delta m^2 < 1\, {\rm eV^2} $}   & {\small  $\theta \sim 2^{\circ}$}    \\ \hline

\end{tabular}
\caption{\small\sf  The fit values of neutrino oscillation parameters from solar, atmospheric, and accelerator experiments.}
\label{tasol}
\end{center}
\end{table}
%

\section{Results and Conclusions}

We have plotted the resonance condition for two- and three-neutrino mixing.  For two-neutrino mixing,  we have used the fit values of neutrino oscillation parameters from solar, atmospheric, and accelerator experiments as shown in Table 1. For three neutrino mixing,   we use the following parameters for this analysis

\begin{figure}[t!] 
\vspace{0.5cm}
{\centering
\resizebox*{0.45\textwidth}{0.3\textheight}
{\includegraphics{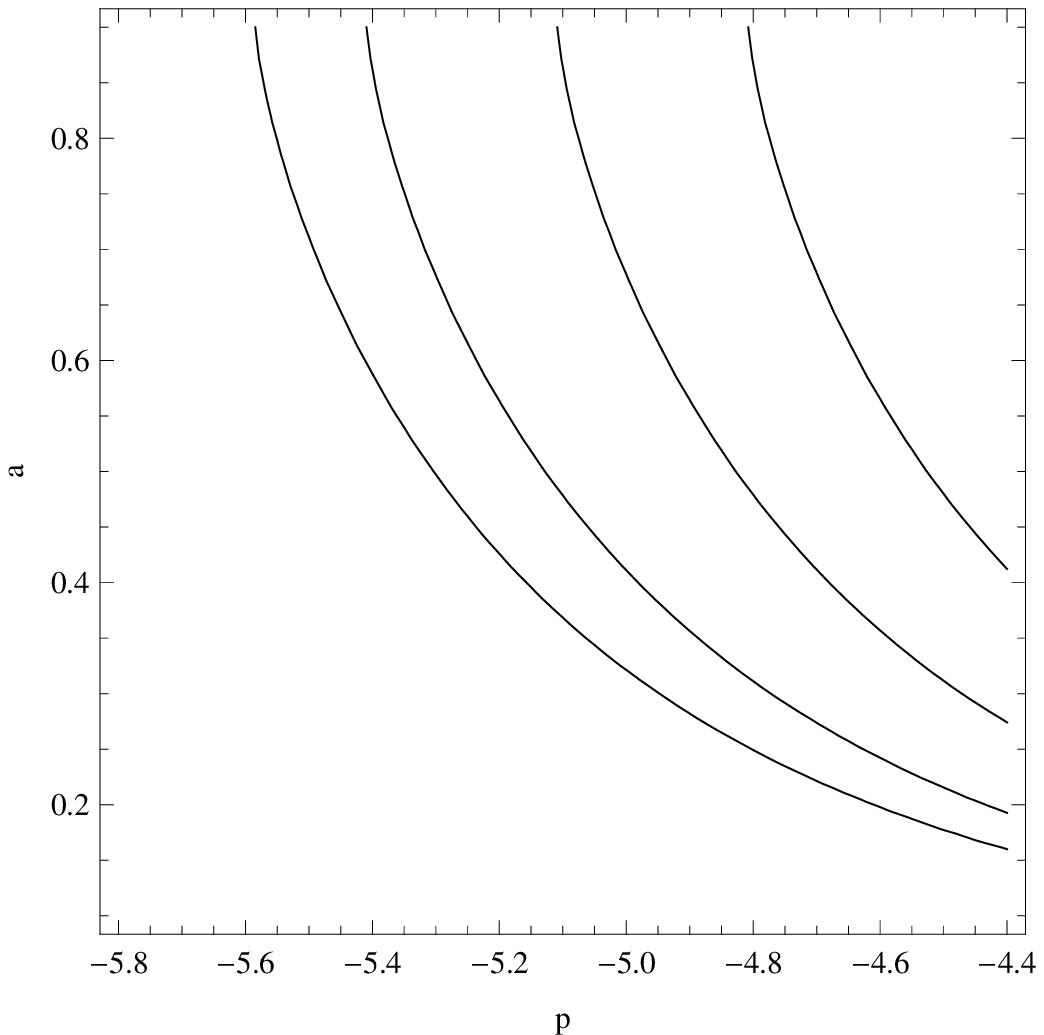}}
\resizebox*{0.45\textwidth}{0.3\textheight}
{\includegraphics{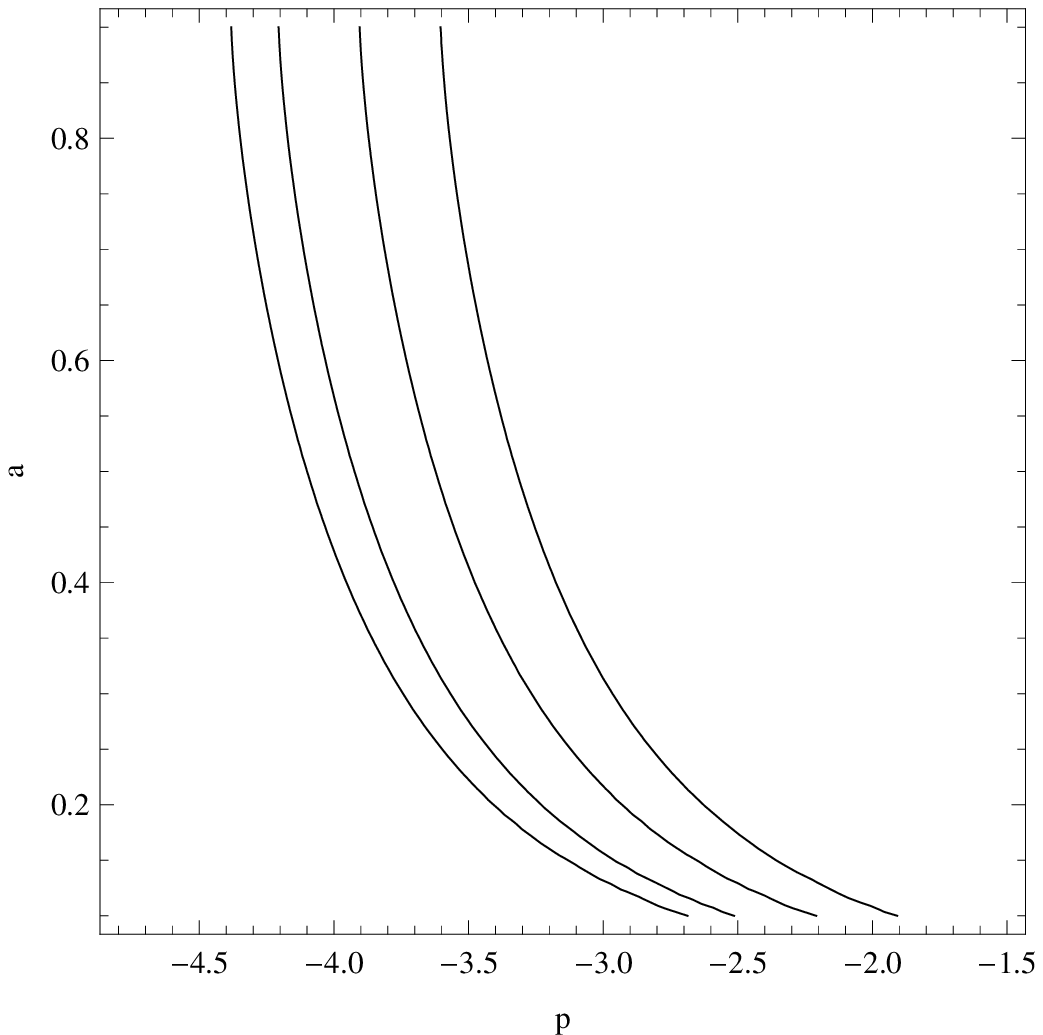}}
\resizebox*{0.45\textwidth}{0.3\textheight}
{\includegraphics{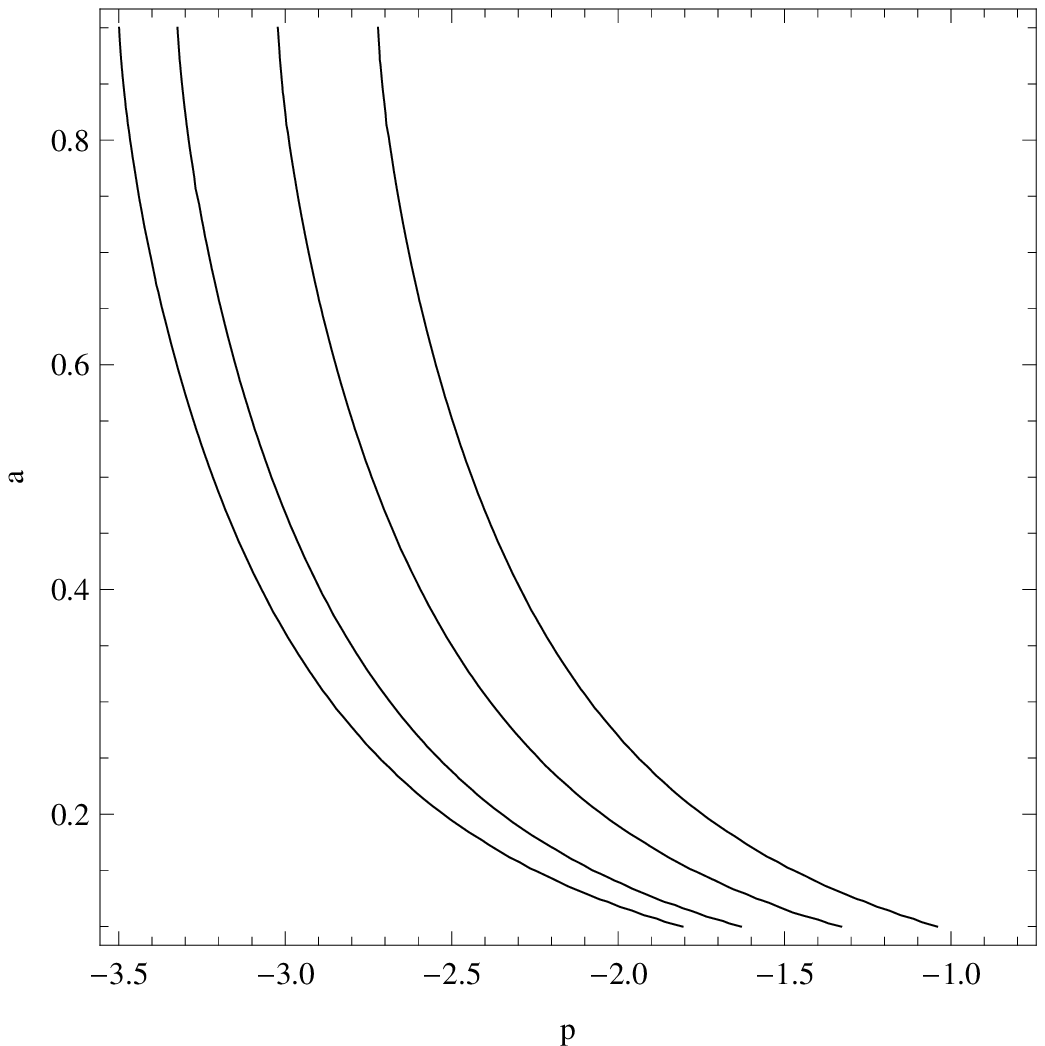}}
\resizebox*{0.45\textwidth}{0.3\textheight}
{\includegraphics{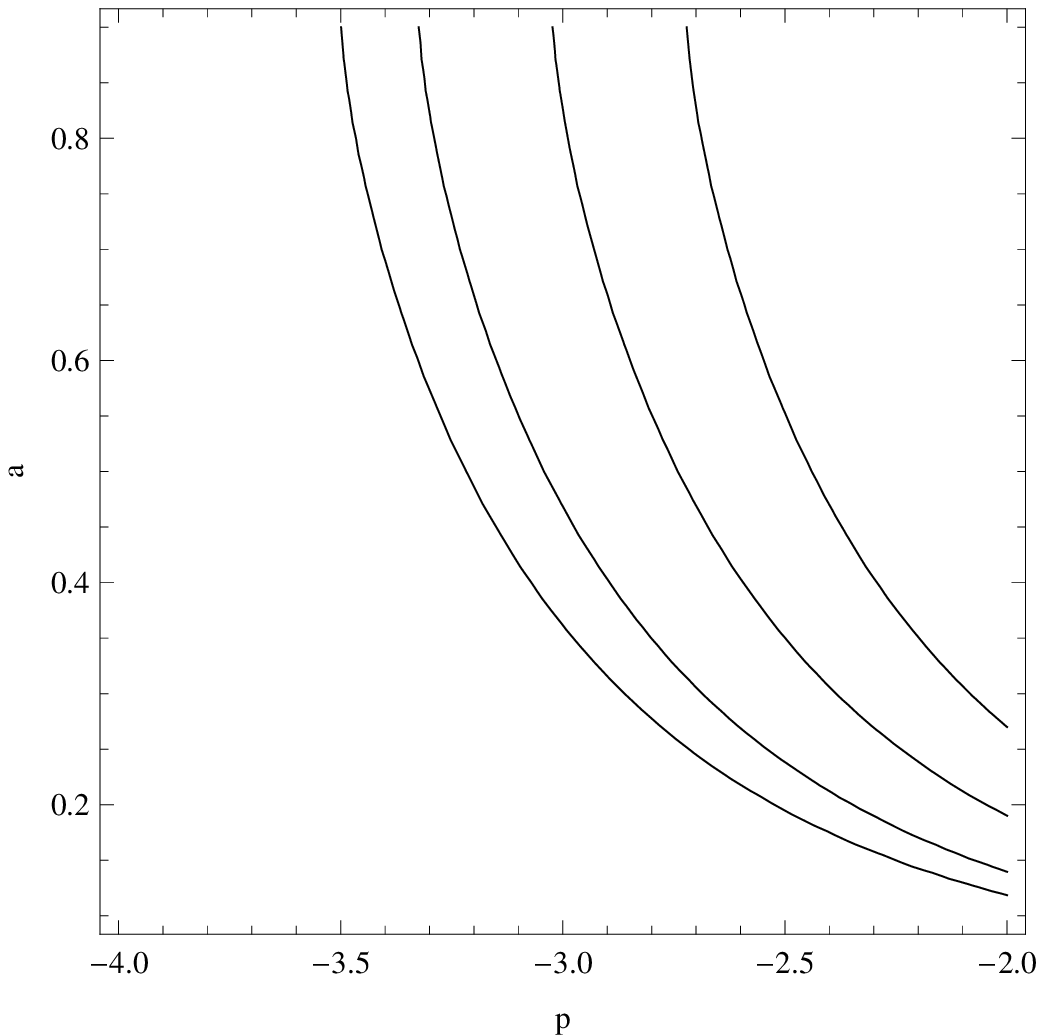}}
}
\caption{ Plot of $a_\star$ vs. chemical potential  ($\mu=m_e10^p$) for which the resonance condition is satisfied (eq. \ref{recond}).  We have used the best-fit parameters of two- (top figure) and three-flavor (bottom figure) neutrino oscillations. For two flavors  we use solar (left) and atmospheric (right) neutrino oscillations and  for three flavors  we use  the large and small  value of   mass difference  $\delta m^2_{32}= 10^{-2.58}$ eV (left) and  $\delta m^2_{32}= 10^{-2.72}$ eV (right), respectively. }
\label{osc_neu}
\end{figure}

\bary
{\rm for}&&\,\,\sin^2_{13} < 0.053: {\rm \cite{aha11}}\cr
&&\Delta m_{21}^2= (7.41^{+0.21}_{-0.19})\times 10^{-5}\,{\rm eV^2};   \hspace{0.1cm} \tan^2\theta_{12}=0.446^{+0.030}_{-0.029},\cr
{\rm for}&&\,\,\sin^2_{13} < 0.04: {\rm \cite{wen10}}\cr
&& \Delta m_{23}^2=(2.1^{+0.5}_{-0.2})\times 10^{-3}\,{\rm eV^2};    \hspace{0.1cm} \sin^2\theta_{23}=0.50^{+0.083}_{-0.093}.
\eary
Analyzing the resonance condition for two-neutrino mixing we found that, unlike using solar and atmospheric parameters (see top fig. \ref{osc_neu}),  the resonance condition is not satisfied for any  value of $a_\star$ and chemical potential, $\mu$, when we use accelerator parameters.  Also, we plot the resonance condition for three-neutrino mixing  (see bottom fig. \ref{osc_neu}), the left column corresponds to  the large value of   mass difference  ($\delta m^2_{32}= 10^{-2.58}$ eV)  while the small value  is used in the right column  ($\delta m^2_{32}= 10^{-2.72}$ eV).\\
Since we are dealing with pop III stars, which produce very massive stellar BHs (a few $100$ to $\sim 1,000 \msun$), and the luminosity depends on both, $B^2$ and $M_h^2$, a collapsar can have a much lower magnetic field (up to 4 orders of magnitude lower than for typical population I or II collapsars).
Hence, in the treatment of pop III collapsars, we can afford to only use sub-critical magnetic fields ($B_c<m^2/e$). 

\Acknowledgements
NF gratefully acknowledges a Luc Binette-Fundaci\'on UNAM Posdoctoral Fellowship.

\end{document}